# Breakdown of the isotropy of diffuse radiation as a consequence of its diffraction at multidimensional regular structures

Vladimir V. Savukov [*]

[*] D. I. Ustinov Voenmekh Baltic State Technical University, St. Petersburg

This article is devoted to the results of an experimental test of the theoretical assumption that the basic axiomatic postulate of statistical physics according to which it is equally probable for a closed system to reside in any of the microstates accessible to it may be invalid for nonergodic cases. In the course of photometric experiments for the purpose of recording the predicted loss of isotropy by a diffuse light field when it came into contact with a two-dimensional phase-type diffraction grating, a significant deviation from Lambert's law was detected when the diffuse photon gas was scattered by the grating surface. This caused angular anisotropy of the radiation fluxes to appear in the initially homogeneous light field. These results provide a basis for revising the determination of the most probable macrostate of a closed system.

*St. Petersburg, Russia, 2009*

## Introduction

As early as 1913, Gaede [1], and later other researchers [2, 3], rigorously proved that a diffuse gas, including a photon gas, can be in a state of comprehensive dynamic equilibrium with a certain reflective surface, provided the angular characteristic of the scattering-probability density, generalized for all the particles of this gas, is described by Lambert's cosine law[1] or Knudsen's cosine law[2] [4, 5].

In other words, for a diffuse photon gas[3] of any spectral composition[4], when its particles undergo "elastic scattering" (without absorption) on any object, Lambert's law is formally required to be satisfied[5]. This requirement is a particular consequence of the zeroth law of thermodynamics.

## Goal of the paper

The goal of this paper is to directly experimentally test the theoretical assumption proposed earlier by the author [6], according to which Lambert's law can break down in certain cases of diffraction scattering of a diffuse (isotropic) photon gas on the surface of a specially organized diffraction optical element (DOE).

## Methodology for carrying out the work

The assumption that it is possible for Lambert's law to break down was tested in the course of photometric experiments for the purpose of recording the possible loss of isotropy by a diffuse light field, predicted as the consequence of its contact with a two-dimensional phase diffraction grating of sinusoidal profile.

---

[1] This name is used in optics when one considers how a flux of photons of electromagnetic radiation interacts with a surface.

[2] This name is ordinarily used in molecular dynamics or in other cases in which the particles of the gas of interest have a nonzero rest mass.

[3] By a diffuse photon gas is here meant unpolarized incoherent electromagnetic radiation for whose individual photons any possible angular orientation of their wave vectors in geometrical space is implemented with equal probability.

[4] The principle of detailed balance assumes that the isotropy properties are conserved for each individual spectral component of the radiation.

[5] At a visual level, this means the following: Totally reflective objects of any macroscopic shape with a matt, mirror, or diffraction-scattering surface located in a closed volume filled with diffuse radiation must not create in this "prepared" isotropic light field any brightness gradients that could be recorded by an external observer.

Each of the experiments was as follows: A DOE with cylindrical surface shape was placed in the central part of the volume of a photometric chamber. A diffuse light field with a spectral composition of the visible range was created inside this chamber. The DOE was photographed on the background of the inner wall of the photometric chamber through a small opening in its housing. The recorded pattern of the assumed distortion of the initially diffuse radiation by the surface of the DOE was processed by the methods of analysis of variance and regression analysis, which detect the presence or absence of the expected anisotropy phenomena.

In order to monitor whether the results obtained here are reproducible, each DOE was independently tested in three different experimental assemblies that differed in the designs of the photometric chambers, the primary sources of radiation, and the methods of rendering it stochastic, as well as the parameters of the recording camera.

## Description of the experimental assemblies and their elements

1. **Experimental assembly No. 1** is based on a cylindrical photometric chamber and an interchangeable outer local source of primary light, which was introduced into the chamber through a massive filter made from acrylic milk plates with a matt surface, arranged on its open end. Such a design does not impose any limitations on the size, type, and power of the radiation source, and this is convenient when varying the different forms of these sources. Five types of lamps, including LED, luminescent, and halogen (incandescent) lamps were used as the local source during the experiments, with various characteristics of the radiation spectra.

   – The shape of the chamber is cylindrical (the position of the symmetry axis is vertical).
   – The (inner) diameter of the chamber is 260 mm.
   – The length along the (outer) symmetry axis is 185 mm.
   – The opening for the camera lens is 5.0 mm in diameter ($\cong 0.00768\%$ of the area of the entire inner surface of the chamber), and the center of the opening for the lens is 85 mm away from the lower surface of the chamber.
   – The coating material of the inner surface of the chamber is bright-white acrylic-base matt paint. It has a spectral reflectance of $\approx 0.97$, averaged over the visible region, and the gloss according to ISO 2813 at 85° is no more than 10%.

2. **Experimental assembly No. 2** is based on a cylindrical chamber and an inner source of stochasticized light, uniformly distributed over the entire inner surface.
   - The shape is cylindrical (the position of the symmetry axis is vertical).
   - The (inner) diameter of the chamber is 220 mm.
   - The length along the (outer) symmetry axis is 155 mm.
   - The opening for the camera lens is 6.5 mm in diameter ($\cong 0.01826\%$ of the area of the entire inner surface of the chamber), and the center of the opening for the lens is 85 mm away from the lower surface of the chamber.
   - The coating of the inner surface of the chamber is a light-scattering underlayer of bright-white matt paint with diffuse reflectance for the visible spectrum of $\approx 0.97$, coated with a layer of a solid colloidal 10% solution of **«Penta L-1»** self-luminescent phosphor in transparent matt lacquer.

   Before beginning each experiment, a special light-accumulating coating of the side and end walls of the chamber (see the description of the inner coating) was irradiated for a long time with a fluorescent "pump" lamp, after which the inner surface of the chamber emitted blue visible light with a dominant wavelength of about $\lambda \approx 420$ nm for about eight hours. During the first 30 min, the brightness of this light was sufficient for any necessary photometric work.

   Despite the inconvenient character of the preparation of each experiment and the fixed spectrum of the resulting light, the isotropy of the radiation formed in chamber No. 2 was virtually ideal. The particles of finely dispersed phosphor introduced into the volume of the transparent matt lacquer acted as an enormous multitude of elementary light sources uniformly distributed over the entire inner surface of the chamber. However, a lower concentration of phosphor particles in the composition of the colloidal "suspension" did not disturb the additional stochastization of the radiation emitted by these particles by the matt-white substrate of the described two-layer coating.

3. **Experimental assembly No. 3** has a spherical chamber and an interior source. This is a chamber that is "classical" in its shape among those that are usually adopted to create diffuse light fields. A 40-W capsule-type baseless **KGM-12 V** miniature incandescent halogen lamp with a continuous "Planck" radiation spectrum and a color temperature of about $T \approx 2700°K$ was used in the assembly. The light source lies directly at the center of the chamber, so that the DOE sample "blocked" this source from the camera.
   - The shape of the chamber is spherical (Fig. 1a).
   - The diameter of the (inner) chamber is 400 mm.

- The diameters of the openings for the camera lens are 32, 16, and 8 mm (respectively 0.16%, 0.04%, and 0.01% of the area of the entire inner surface of the chamber).
- The coating material of the inner surface of the chamber is similar to the inner coating of the chamber in assembly No. 1.

## Samples of diffraction optical elements

Studies were carried out on six samples of cylindrical shape whose scattering surfaces were DOEs fabricated in the form of flexible replicas of a diffraction grating (Fig. 1b). The working part of each DOE is square and consists of equal rectangular strips with an aluminum reflective coating. Microrelief with specified parameters (diffraction sections acting as a phase grating) is formed on half of these strips, while the strips separating them possess a mirror-smooth surface.

Because the DOEs consist of alternating (adjacent) diffraction and mirror sections, it becomes objectively possible to monitor the homogeneity of the initial light field. Actually, such adjacent sections of the DOEs simultaneously are under identical illumination conditions. Therefore, the absence of initial isotropy of the light field would become obvious in the character of its reflection from the mirror sections.

The geometry of the microrelief of the diffraction sections of the DOEs has a quasi-sinusoidal profile, characterized by proportions between the step $S_o$ (the period) and the amplitude variation A of the height that are standard for a sinusoidal function. The step of the grating is $S_o \approx 417, 833, 1667,$ and $10000\,\text{nm}$, and the peak-to-peak amplitude is $A = \pm S_o/2\pi$.

The dimensionality of the microrelief geometry of the diffraction sections of the individual DOEs was either one-dimensional (linear) or two-dimensional (an orthogonal crossed grating with the rulings intersecting at an angle of 90°). The ruling lines of the different diffraction gratings were oriented (sloped) either at an angle of 45° or at an angle of 90° with respect to the sides of the square working region of the DOEs. Each DOE is attached to a massive metal cylinder with a mirror side surface and a matt-white end surface (the reference cylinder) so that the diffraction and mirror strips of the DOE were horizontal when the cylinder was vertically positioned.

## Recording apparatus

Experimental assemblies Nos. 1 and 2 (with cylindrical chambers) included an ultracompact **Canon™ Digital Ixus 90 IS** camera, and assembly No. 3 (with a spherical chamber) included a compact **Canon™ digital PowerShot S2 IS** camera.

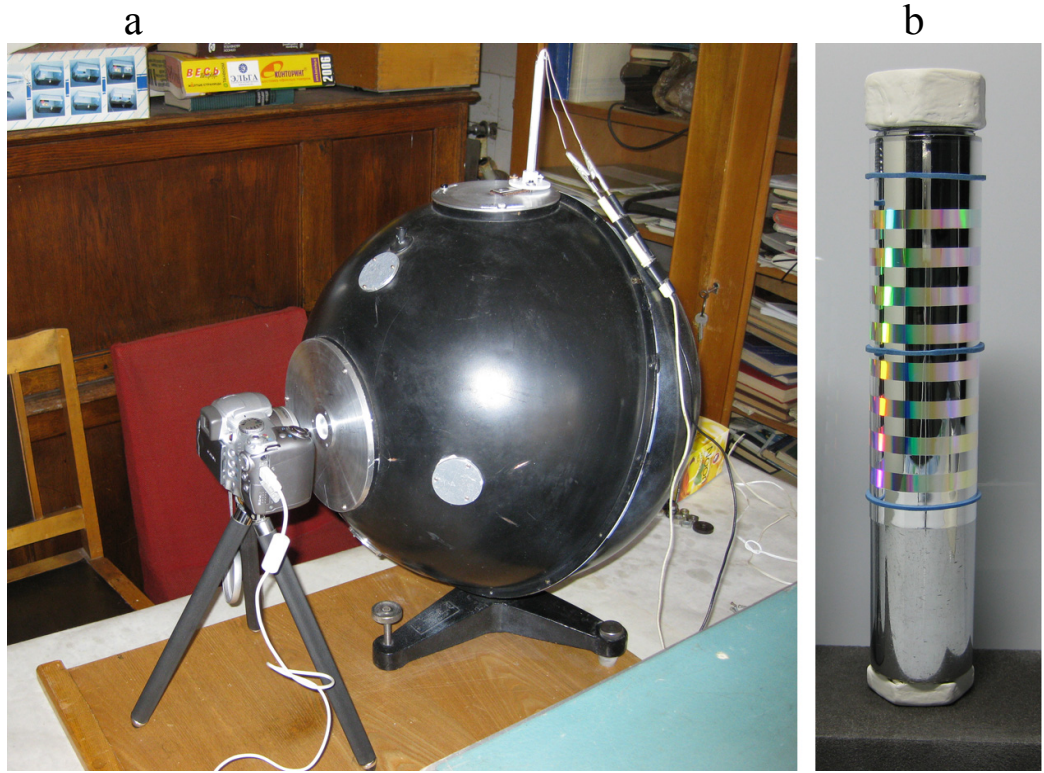

Fig. 1. (a) Installation based on photometric chamber No. 3, (b) external view of diffraction optical element.

## Experimental results

Our experiments revealed a significant deviation from Lambert's law when a diffuse photon gas is diffracted on the surface of a two-dimensional diffraction grating with a quasisinusoidal profile of the microrelief. This deviation, in turn, caused angular anisotropy of the radiation fluxes in an initially homogeneous light field.

Figure 2 shows two photographs that illustrate the manifestation of the anisotropic scattering effect. In both cases, the working surfaces of the DOEs are flexible replicas that have alternating horizontal sections of mirror type (shown lighter on the photographs) and sections with two-dimensional quasi-sinusoidal microrelief, the direction of whose rulings makes 45° angles with the directrices of the reference cylinders. Both samples are successively placed in a photometric chamber filled with stochasticized light. A fluorescent lamp with a three-band spectrum was used as the primary radiation source, whose working wavelengths were 435±10 nm, 545±10 nm, and 610±10 nm. The color temperature of the source was $T \approx 4200°K$, while the colorrendition coefficient was Ra ≈ 80.

The photograph (Fig. 2a) shows the image of a DOE whose grating step is too large (10 μm) for the predicted anisotropy effects to manifest themselves in the visible region. Figure 2b shows a picture of a DOE with a grating step of

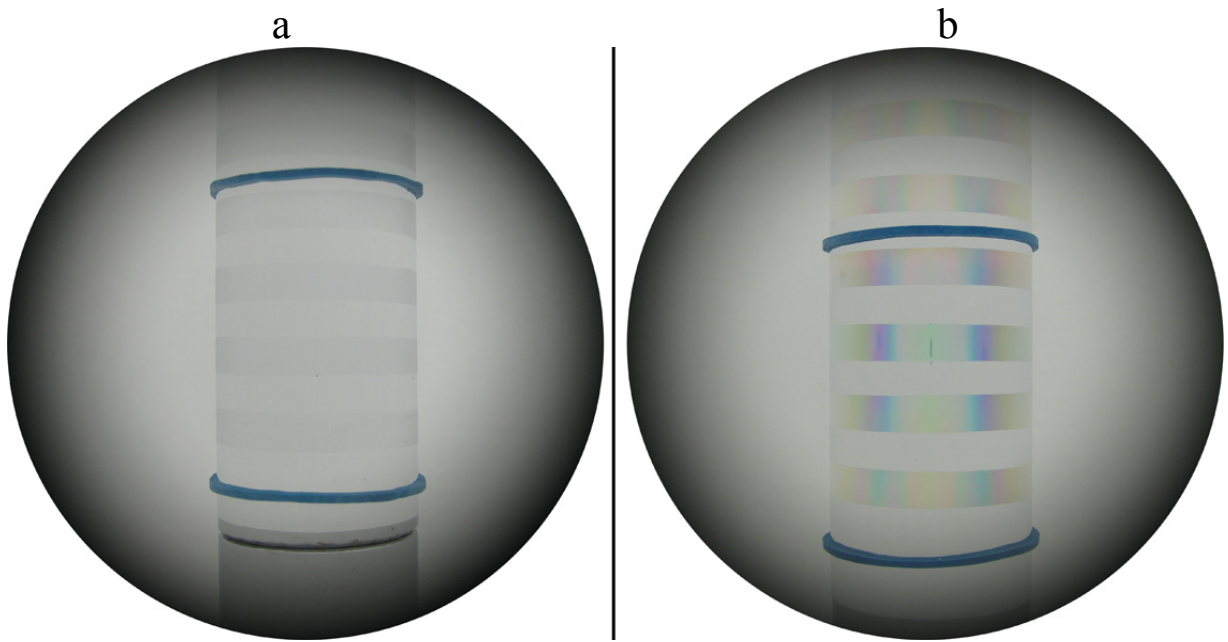

Fig. 2. Photographs of DOEs with two-dimensional quasi-sinusoidal profile of the microrelief.
(a) with grating step 10 μm, (b) with step 833 nm.

The photograph (Fig. 2a) shows the image of a DOE whose grating step is too large (10 μm) for the predicted anisotropy effects to manifest themselves in the visible region. Figure 2b shows a picture of a DOE with a grating step of 833 nm. The local decrease of the spectral brightness, clearly observable on the diffraction sections of its side surface, in this case is about 30% of the mean overall level. Here the angular anisotropy of the brightness of the initial diffuse light field does not exceed 2%.

The reproducibility of the detected effect was tested for a large number of structurally feasible sets of DOEs (six versions), photometric chambers (three versions), and radiation sources (eight versions). The results make it possible to conclude that the observed losses of isotropy by the initially homogeneous diffuse radiation has an objective character [7].

It is theoretically shown in Ref. [7] that the dip in the brightness level of the DOE surface must be observed predominantly in planes oriented at a tangential angle of 45° to the ruling lines of a two-dimensional phase diffraction grating, while the effect must be absent on one-dimensional gratings. The optimum value of the reflection angle β, measured from the normal to the macrosurface of the grating, in this case is:

$$\beta \approx \arcsin(1 - \lambda/(\sqrt{2}S_o)) \qquad (1)$$

where $\beta$ – is the optimum angle for observing the surface;

$S_o$ – is the grating step in a direction orthogonal to the rulings;

$\lambda$ – is the dominant wavelength of the scattered radiation, with $\lambda \leq \sqrt{2}S_o$

It was confirmed that these predictions are valid with high accuracy in all the experiments that were carried out. The wavelength dependence of the angular localization of the effect is seen in Fig. 2b, where those spectral components of the initial metameric white light are observed that are complementary (paired) with respect to the "suppressed" components. The effect acquires the highest intensity if the dominant wavelength becomes equal to $\lambda \approx \sqrt{2}S_o$.

It is important to point out that the diffraction-scattering mechanism considered in Ref. [7] is partly analogous to that implemented in forming the Rayleigh-Wood anomalies [8, 9], and this is reflected in the dependence shown in Eq. (1).

## Conclusions

The statistical physics of equilibrium systems is based on the postulate that all the microstates accessible to a closed system are equally probable. This postulate has the character of a hypothesis declared a priori, the adoption of which makes it possible to assert that there can be only one equilibrium macrostate in the test system, with the parameters of this single state being rigorously fixed. The "zeroth law" serves as a functional analog in thermodynamics.

The detected mechanism by which a light field is scattered at the surface of a DOE creates in the phase space accessible to a physical system local regions with nonzero divergence of the flux of the phase trajectories of the photons (sources and sinks). This makes such a system nonergodic[6], and hence incompatible with the axiomatics of statistical physics.

The resulting effect gives an objective basis for revising the existing concept of the most probable macroscopic state of a closed system, since it is obvious that such a state may not correspond to the definition of thermodynamic equilibrium.

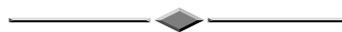

---

[6] The property of ergodicity assumes the validity of the microcanonical hypothesis of statistical physics that the results of averaging over time and over phase are identical when the values of the macroscopic parameters of a system are being computed.

# References

1. *Gaede W.* The external friction of gases // Annalen der Physik, 1913, B. **41**, 289-336.

2. *Epstein P. S.* On the Resistance Experienced by Spheres in their Motion through Gases // Physical Review, 1924, v. **23**, 710-733.

3. *Clausing P.* Cosine law of reflection as a result of the second main theorem of thermodynamics // Annalen der Physik, 1930, B. **4**, 533-566.

4. *Goodman F. O., Wachman H. Y.* Dynamics of Gas-Surface Scattering // Academic Press, New York, 1976, 423 p.

5. *Ramsey N. F.* Molecular Beams // Clarendon Press, Oxford, 1956, 466 p.

6. *V.V. Savukov* Refining the axiomatic principles of statistical physics. // Deposited at the All-Union Institute of Scientific and Technical Information, No. 1249-B2004, – 177 p.[7]

7. *V.V. Savukov* Breakdown of Lambert's law for diffraction of a diffuse photon gas at multidimensional regular structures. // Deposited at the All-Union Institute of Scientific and Technical Information, No. 507-B2009, – 49 p.[8]

8. *Wood R.W.* On a remarkable case of uneven distribution of light in a diffraction grating spectrum // Philosophical Magazine, 1902, v. **4**, 396-402.

9. *Lord Rayleigh.* On the Dynamical Theory of Gratings // Proceedings of the Royal Society of London, 1907, Series A 79, 399-416.

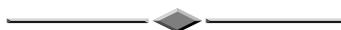

---

[7] URL: http://www.savukov.ru/viniti_1249_b2004_full_rus.pdf

[8] URL: http://www.savukov.ru/viniti_0507_b2009_full_rus.pdf